\documentstyle[epsfig]{aipproc}
\begin{document}
\begin{flushright}
 UR-1610 \\
 ER/40685/949
\end{flushright}
\title{Vector Boson Production at Hadron Colliders: Results
from HERWIG and Resummed Calculations\footnote{Talk given at
the 22nd annual MRST (Montreal-Rochester-Syracuse-Toronto)
Meeting on High Energy Physics
(MRST 2000), Rochester, NY, U.\ S.\ A., 8-9 May 2000.}}
\author{G. Corcella}
\address{Department of Physics and Astronomy, University of Rochester,\\
Rochester, NY 14627, U.\ S.\ A.}
\maketitle
\begin{abstract}
We discuss vector boson
production at hadron colliders and the treatment of the 
initial-state radiation according to 
Monte Carlo parton shower simulations and resummed calculations.
In particular, we investigate 
the effect of matrix-element corrections to the HERWIG event generator
on $W/Z$ transverse momentum distributions.
\end{abstract}
\section*{Introduction}
The production of vector bosons $W$, $Z$ and $\gamma$ \cite{altarelli} is 
one of the most interesting processes in the phenomenology of hadron 
collisions and provides an environment to test both Quantum Chromodynamics and
the Standard Model of electroweak interactions (see \cite{lhc,lh} for a 
review).
The lowest-order processes $q\bar q'\to V$ are
not sufficient to make reliable predictions, but the initial-state 
radiation should be taken into account.
Monte Carlo event generators and resummed analytical calculations are 
available tools to describe the multiple radiation accompanying
the incoming hadrons.

Standard Monte Carlo algorithms \cite{herwig,pythia}
describe the initial-state parton showers
in the soft/collinear approximation, but can have `dead zones', 
where no radiation is permitted.
The radiation in these regions 
can be generated by the use of the exact first-order
matrix element. Referring to the HERWIG event generator, 
matrix-element corrections to Drell--Yan processes have been implemented in
\cite{corsey}, following the general method of \cite{sey},
and included in the new version HERWIG 6.1 \cite{herwig61}. 

Another possible approach 
consists of performing an analytical resummation of the
large logarithmic coefficients which multiply the strong coupling 
constant. Considering the transverse momentum $q_T$ distribution, logarithms
of the ratio $m_V/q_T$, $m_V$ being the vector boson mass, arise in
calculating higher-order corrections to the Born process.
The resummation of these logarithms, which are large at 
small $q_T$, was initially 
proposed by Dokshitzer, Dyakonov and Troyan (DDT) \cite{ddt}, 
then accomplished by Collins, Sterman and Soper (CSS) \cite{css}.
CSS performed the resummation in the
space of the impact parameter $b$, which is the Fourier conjugate of $q_T$.
Their results have been implemented numerically in \cite{ly,balazs},
while more recent analyses can be found in \cite{ellis,fnr,kulesza}, 
where the resummation is performed in both $q_T$- and $b$-space.

In this paper, we review some results for the $W/Z$ transverse momentum 
distribution according to the HERWIG event generator and resummed 
calculations.\footnote {See also \cite{huston} for a similar comparison 
for Higgs production at hadron colliders.}

\section*{The HERWIG parton shower algorithm}
HERWIG simulates the initial-state radiation in
hadron collisions according to a `backward evolution' \cite{marweb}, 
in which the scale
is reduced away from the hard vertex and traces the hard-scattering partons
back into the incoming hadrons. 
The branching algorithm relies on the universal structure of the elementary 
probability in the leading infrared approximation.
The probability of the emission
of an additional soft/collinear 
parton from a parton $i$ is given by the general result:
\begin{equation}
  \label{elementary}
  dP={{dq_i^2}\over{q_i^2}}\;
  {{\alpha_S\left(\frac{1-z_i}{z_i}q_i\right)}\over {2\pi}}\;
  P_{ab}(z_i)\; dz_i\;
  {{\Delta_{S,a}(q^2_{i\mathrm{max}},q_c^2)}\over{\Delta_{S,a}(q_i^2,q_c^2)}}\;
{{x_i/z_i}\over x_i}\;{{f_b(x_i/z_i,q_i^2)}\over {f_a(x_i,q_i^2)}}.
\end{equation}
The ordering variable is
$q_i^2=E^2\xi_i$, where $E$ is the energy of the parton that splits
and $\xi_i={{p\cdot p_i}\over{E E_i}}$, with $p$ and $p_i$ being the
four-momenta of the splitting and of the emitted parton respectively;
$z_i$ is the energy fraction of the outgoing space-like
parton with respect to the incoming one; $P_{ab}(z)$ is the Altarelli--Parisi
splitting function for a parton $a$ evolving in $b$.
In the approximation of massless partons, we have
$\xi_i=1-\cos\theta$, where $\theta$ is the emission angle to the
incoming hadron direction. 
For soft emission ($E_i\ll E$),  ordering according to $q_i^2$
corresponds to angular ordering. 
When the emission is hard, the energy of
the radiated parton is similar to that of the splitting parton, so
$q_i^2$-ordering corresponds to transverse momentum ordering.
In (\ref{elementary}) $f_a(x_i,q_i^2)$ is the parton distribution function
for the partons of type $a$ in the initial-state hadron, $x_i$ being the
parton energy fraction.
The function 
\begin{equation}
\Delta_S(q_2^2,q_1^2)=\exp\left[-{{\alpha_S}\over {2\pi}}
\int_{q_1^2}^{q_2^2}{{dk^2}\over{k^2}}\int_{Q_1/Q_2}^{1-Q_1/Q_2}
{dzP(z)}\right] \end{equation}
is the Sudakov form factor, expressing the probability of no-resolvable 
branching in the range $q_1^2<q^2<q_2^2$.
The ratio of form factors
in (\ref{elementary}) is therefore the probability of no branching at higher 
values of $q_i^2$. Unitarity dictates that the Sudakov form factor
sums up all-order virtual and unresolved contributions. 
In (\ref{elementary}), $q_{i\mathrm{max}}$ is the maximum value of $q$, fixed
by the hard process, and $q_c$ is the value 
at which the backward evolution is terminated, corresponding, in the case of 
HERWIG, to a cutoff on the transverse momentum
of the showering partons. 
However, since $q_c$ is smaller than the minimum scale at which 
the parton distribution functions are evaluated, an additional cutoff 
on the evolution variable $q_i^2$ has to be set.

If the backward evolution has not resulted in a
valence quark, an additional non-perturbative parton emission is generated to
evolve back to a valence quark. Such a valence quark has a Gaussian
distribution with respect to the non-perturbative intrinsic transverse momentum
in the hadron, with a width $q_{T{\mathrm{int}}}$ 
that is an adjustable parameter and whose default value is zero.

We need finally to specify
the showering frame, the variables $q_i^2$ and $z_i$ being 
frame-dependent. One can show that, as a result of the 
$q^2$-ordering, the maximum $q$-values of two colour connected partons
$i$ and $j$ are related via $q_{i\mathrm{max}}q_{j\mathrm{max}}=p_i\cdot p_j$,
which is Lorentz-invariant.
For vector boson production, symmetric limits are set in HERWIG:
$q_{i\mathrm{max}}=q_{j\mathrm{max}}=\sqrt{p_i\cdot p_j}$.
Furthermore, the energy of
the parton which initiates the cascade is given by
$E=q_{\mathrm{max}}=\sqrt{p_i\cdot p_j}$.
It follows that ordering according to $q^2$ implies $\xi<z^2$.

The region $\xi>z^2$ is therefore a `dead zone' for the shower 
evolution.
In such a zone the physical radiation is not logarithmically
enhanced, but not completely absent as happens in the 
standard algorithm.
We therefore need to improve the HERWIG parton showers by the use
of matrix-element corrections.
\section*{Matrix-element corrections}
According to \cite{sey},
we populate the `dead zone' of the phase space using the exact
${\cal O}(\alpha_S)$ matrix element (hard correction). 
We also correct the emission in the already-populated region using
the first-order result any time an emission is capable of being the 
`hardest so far' (soft correction), where the hardness of an emission 
is measured in terms of the transverse momentum of the emitted parton
relative to the splitting one.

We consider the
process $q(p_1)\bar q'(p_2)\to V(q)g(p_3)$, define the Mandelstam
variables $\hat s=(p_1+p_2)^2$,
$\hat t=(p_1-p_3)^2$ and $\hat u=(p_2-p_3)^2$ and obtain the total 
phase-space limits 
\begin{eqnarray}
  \hspace{-2cm}
   m_V^2 \;\;<&\hat s&<\;\; s,
  \\\hspace{-2cm}
  m_V^2-\hat s \;\;<&\hat t&<\;\; 0,
\end{eqnarray}
$s$ being the total centre-of-mass energy.
We observe that the soft singularity corresponds to $s=m_V^2$ and
the lines $\hat t=0$ and $\hat t=m_V^2-\hat s$ to collinear gluon
emission.

\begin{figure} 
\centerline{\epsfig{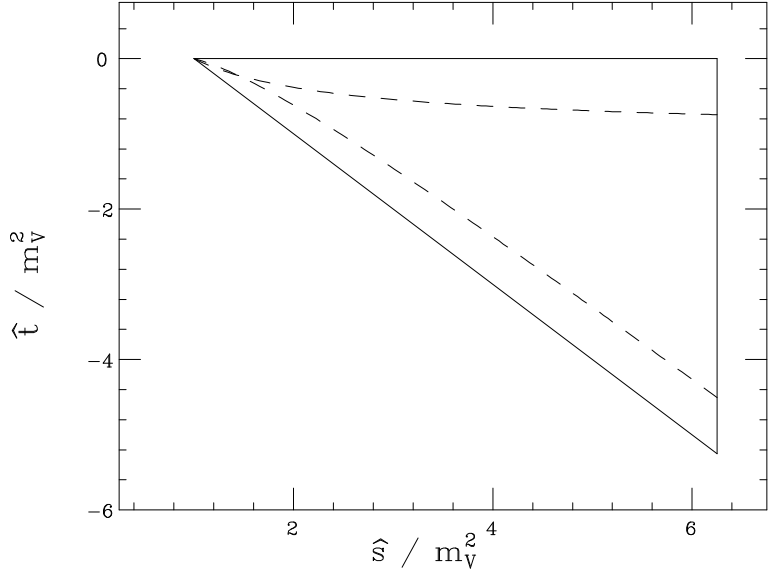}}
\vspace{10pt}
\caption{Total (solid line) and HERWIG (dashed) phase space limits
for $\sqrt{s}=200$~GeV and $m_V=80$~GeV.}
\label{fig1}
\end{figure}

After relating the parton shower variables $z$ and $\xi$ 
to $\hat s$ and $\hat t$, as done in \cite{corsey}, and setting
$\xi<z^2$, one can get the HERWIG phase-space limits in terms of $\hat s$ and 
$\hat t$. 
In Fig.~\ref{fig1} we plot the total and the HERWIG phase space for 
$\sqrt{s}=200$~GeV and $m_V=80$~GeV; the soft and collinear 
singularity are inside the HERWIG region. 

We calculate the differential cross section with respect to $\hat s$ and 
$\hat t$ following the prescription of \cite{sey2}, where it is shown that,
assuming that the rapidity and the virtuality of the vector boson
are fixed by the Born process, the following factorization formula holds:
\begin{equation}
\label{dsigma}
d^2\sigma = \sigma_0{{f_{q/1}(\chi_1)f_{\bar q'/2}(\chi_2)}\over
{f_{q/1}(\eta_1)f_{\bar q'/2}(\eta_2)}}{{C_F\;\alpha_S}\over {2\pi}}
{{d\hat s\; d\hat t}\over{\hat s^2\hat t\hat u}}
\left[(m_V^2-\hat u)^2+(m_V^2-\hat t)^2\right],
\end{equation}
where $f_{q/1}(\chi_1)$ and $f_{\bar q'/2}(\chi_2)$ are the
parton distribution functions of the scattering partons inside the incoming
hadrons 1 and 2 for energy fractions $\chi_1$ and $\chi_2$ in the process
$q\bar q'\to V g$, while $f_{q/1}(\eta_1)$ and $f_{\bar q'/2}(\eta_2)$ refer
to the Born process.

A similar treatment holds for the Compton process 
$q(p_1) g(p_3)\to q'(p_2) V(q)$, as discussed in \cite{corsey}.

The distribution (\ref{dsigma}) or the equivalent one for the Compton process
is implemented 
to generate events in the missing phase space and in the populated 
region every time an emission is the hardest so far.
\section*{Transverse momentum distributions: HERWIG results}
\begin{figure} 
\centerline{\epsfig{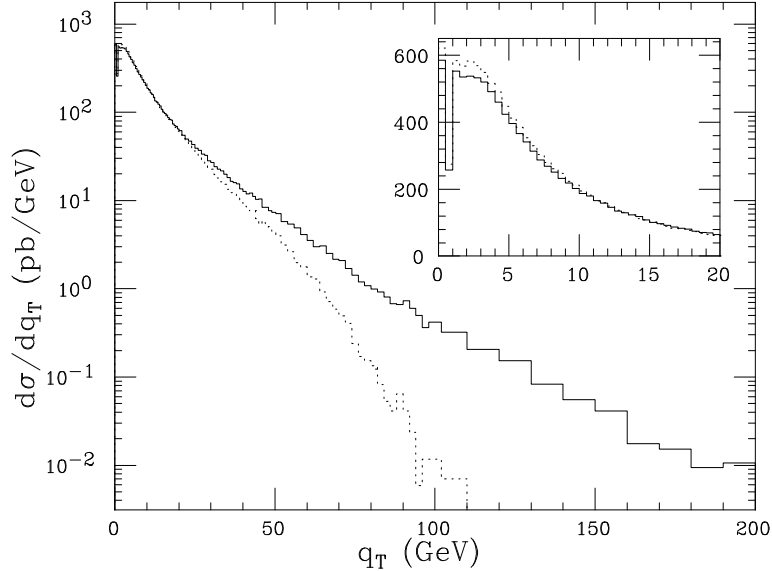}}
\vspace{10pt}
\caption{$W$ transverse momentum distributions at the Tevatron, according
to HERWIG 6.1 (solid line) and HERWIG 5.9 (dotted).}
\label{fig2}
\end{figure}

\begin{figure} 
\centerline{\epsfig{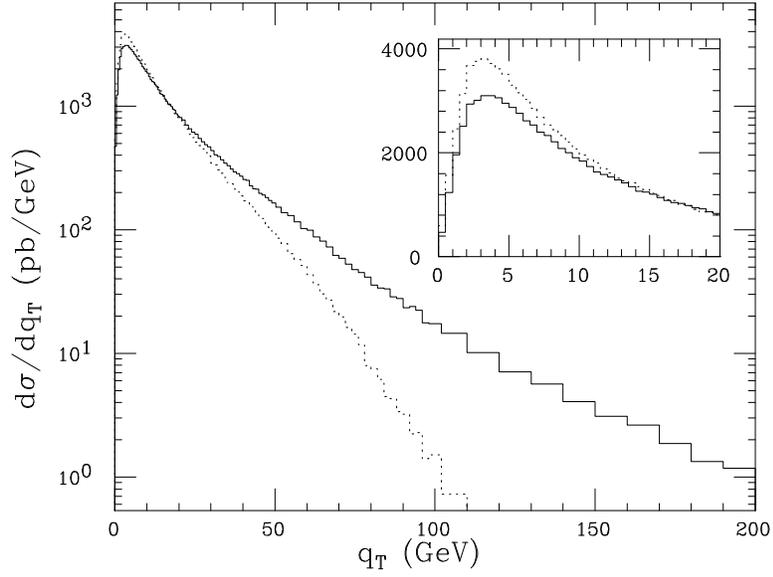}}
\vspace{10pt}
\caption{As in Fig.~\ref{fig2}, but at the LHC.}
\label{fig3}
\end{figure}

An interesting observable to study 
is the vector boson transverse momentum, which
is constrained to be $q_T<m_V$ in the soft/collinear approximation.
After matrix-element corrections, a fraction of events at 
higher $q_T$ is to be expected. 
In Figs.~\ref{fig2} and \ref{fig3} we plot the $W$ $q_T$ distribution at the
Tevatron and at the LHC, according to HERWIG 5.9 and HERWIG 6.1, the new
version including matrix-element corrections, for 
$q_{T{\mathrm{int}}}=0$.
We see a big effect at large $q_T$: after some $q_T$ the 5.9 version
does not generate events anymore, while we still have a non-zero
cross section after matrix-element corrections.

Moreover, a slight suppression can be seen at small $q_T$. 
It is related to the fact that, although we are providing the 
shower with the tree-level ${\cal O}(\alpha_S)$ corrections, virtual 
contributions are missing and we still get the leading-order cross 
section. The enhancement at large $q_T$ is therefore compensated by a 
suppression in the low-$q_T$ range.

\begin{figure} 
\centerline{\epsfig{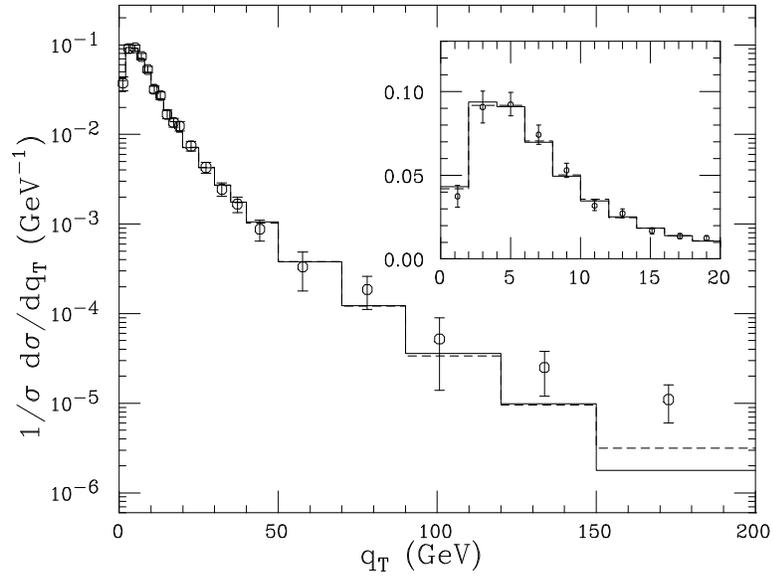}}
\vspace{10pt}
\caption{Comparison of the D\O\ data to the HERWIG 6.1 result,
for an intrinsic $q_T$ of 0 (solid line) and 1 GeV (dashed).}
\label{fig4}
\end{figure}

\begin{figure} 
\centerline{\epsfig{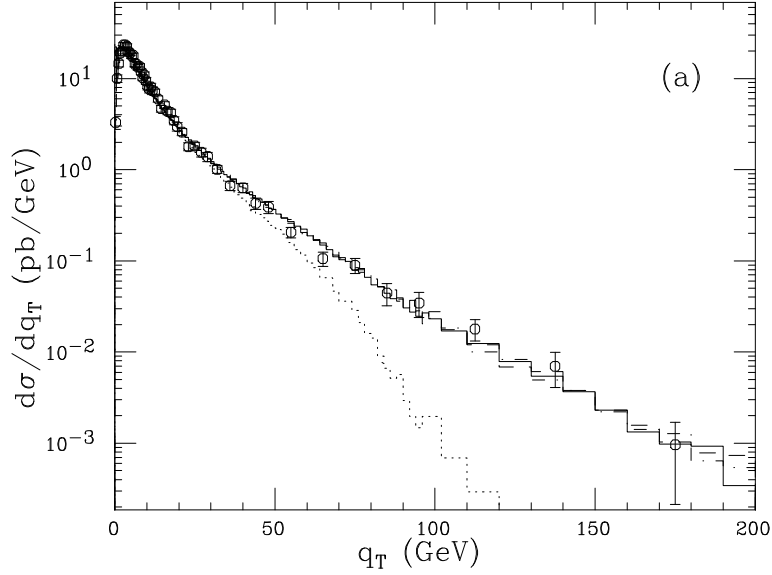}}
\vspace{10pt}
\caption{Comparison of the CDF data on the $Z$ transverse momentum 
to HERWIG 5.9 (dotted line)
and 6.1 for an intrinsic $q_T$ of 0 (solid line), 1 GeV (dashed)
and 2 GeV (dot-dashed).}
\label{fig5}
\end{figure}

\begin{figure} 
\centerline{\epsfig{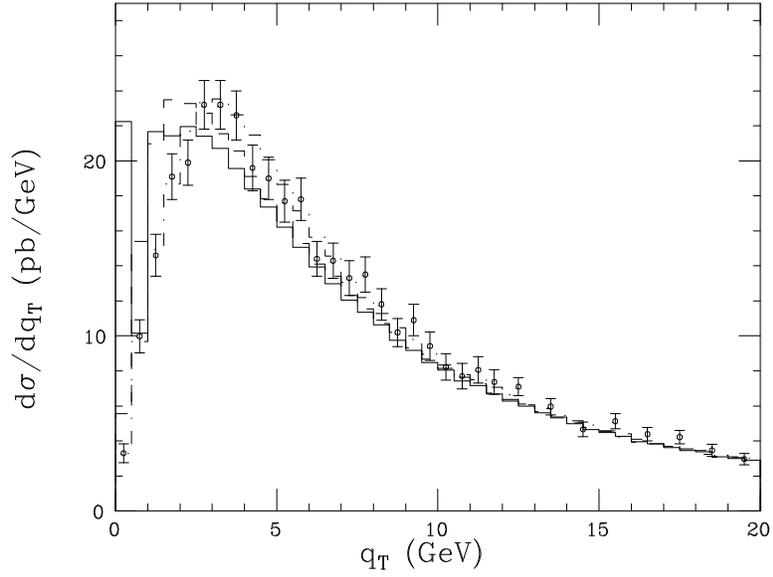}}
\vspace{10pt}
\caption{As in Fig.~\ref{fig5}, but in the low-$q_T$ range.}
\label{fig6}
\end{figure}

\begin{figure} 
\centerline{\epsfig{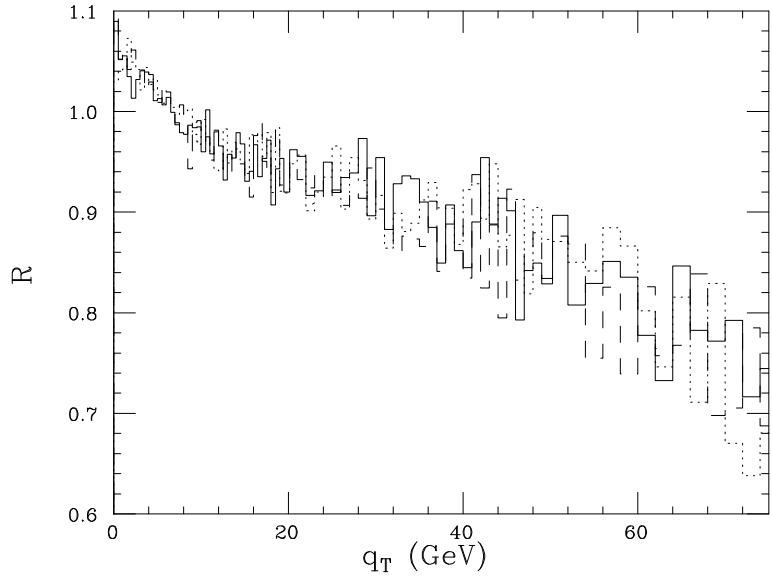}}
\vspace{10pt}
\caption{The ratio of the $W$ to the $Z$ spectrum at small $q_T$ for
an intrinsic transverse momentum of zero (solid), 1 (dashed) and 2 GeV
(dotted).}
\label{fig7}
\end{figure}

It is now interesting to compare the HERWIG results with some Tevatron data.
In Fig.~\ref{fig4} we compare the HERWIG 6.1 distribution with some D\O\ data
\cite{d0} and find reasonable agreement over the whole $q_T$ range.
As shown in \cite{corsey}, smearing the HERWIG 
curve to account for detector effects must be included to achieve this 
agreement. Also, we do not see
any relevant impact of setting $q_{T{\mathrm{int}}}=1$~GeV after detector 
corrections.
In Fig.~\ref{fig5} we compare the HERWIG 5.9 and 6.1 results, for 
different values of $q_{T{\mathrm{int}}}$, with some CDF data \cite{cdf},
already corrected for detector effects. 
We find good agreement after matrix-element corrections, while the 5.9
version is not able to fit in with the data for $q_T>50$~GeV. 
At low $q_T$, the best agreement to the data is obtained for 
$q_{T{\mathrm{int}}}=2$~GeV, as shown in Fig.~\ref{fig6}.
While the $Z$ distribution strongly depends on $q_{T{\mathrm{int}}}$
at small $q_T$, in \cite{corsey1} and Fig.~\ref{fig7} it is shown that the
ratio $R$ of the $W$ and $Z$ spectra is approximately independent 
of it.\footnote{The negative slopes of the plots in Fig.~7 are due
to the $W/Z$ mass difference.}.
Such a ratio is one of the main inputs for the $W$ mass measurement in
hadron collisions and it is good news that it does not depend on unknown
non-perturbative parameters. 
\section*{Resummed calculations}
Another possible approach to study the vector boson transverse momentum 
distribution consists of resumming the logarithmic terms 
$l=\log (m_V^2/q_T^2)$ in the low-$q_T$ range.
It is interesting to compare the HERWIG phenomenological results with those
of some resummed calculations, in particular \cite{ellis} and
\cite{fnr}.
According to \cite{ddt}, the resummed 
differential cross section for $W$ production can be written as:
\begin{eqnarray}\label{ddtcs}
{{d^2\sigma}\over{dm_V^2dq_T^2}} &=& \sigma_0 {d\over{dq_T^2}}
\sum_{q,q'}|V_{q\bar q'}|^2\int_0^1{dx_1 dx_2\; \delta(x_1x_2-\tau)}\nonumber\\
&\times& \left[ f_{q/1}(x_1,q_T)f_{\bar q'/2}(x_2,q_T)\exp[S(m_V,q_T)]+
(q\leftrightarrow\bar q')\right]\;,
\end{eqnarray}
where $V_{q\bar q'}$ is the relevant Cabibbo--Kobayashi--Maskawa matrix
element and $\tau=m_V^2/s$. 
In (\ref{ddtcs}), $\exp[S(m_V,q_T)]$ is a Sudakov-like form factor 
which resums the large logarithms 
associated to the initial-state radiation. It reads:
\begin{equation}
S(m_V,q_T)=-\int_{q_T^2}^{m_V^2}{{{d\mu^2}\over{\mu^2}}
\left[A(\alpha_S(\mu^2))\log{{m_V^2}\over{\mu^2}}+B(\alpha_S(\mu^2))\right]}\;,
\end{equation}
where $A(\alpha_S)$ and $B(\alpha_S)$ can be expanded as:
\begin{equation}\label{abexp}
A(\alpha_S)=A_1\alpha_S+A_2\alpha_S^2+\dots\ ;\ 
B(\alpha_S)=B_1\alpha_S+B_2\alpha_S^2+\dots
\end{equation}
As far as the logarithms which contribute to the resummation are concerned, 
two conflicting nomenclatures exist.
One consists of looking at Sudakov exponent, where the leading logarithms
(LL) are $\sim \alpha_S^nl^{n+1}$ and the next-to-leading ones (NLL)
$\sim\alpha_S^nl^n$. It is straightforward to show that 
the LL contributions are 
obtained by keeping only the $A_1$ term in the expansions (\ref{abexp}) 
while NLL accuracy is achieved by considering 
$A_2$ and $B_1$ as well. In this sense, the approach \cite{fnr}
is NLL.

Another classification relies on the expansion of the exponent
\begin{equation}
S(m_V,q_T)=\sum_n c_{n,n+1}\alpha_S^nl^{n+1}+\sum_n c_{n,n} \alpha_S^nl^n,
\end{equation}
where the leading term is $\sim \alpha_Sl^2$, so that the 
leading contributions to $\exp [S(m_V,q_T)]$ 
are $\sim\alpha_S^n l^{2n}$,
terms $\sim\alpha_S^n l^{2n-1}$ being next-to-leading.
This is equivalent to saying that in the differential cross section
the LL and NLL contributions are 
$\sim (1/q_T^2)\alpha_S^n l^{2n-1}$ and $\sim (1/q_T^2)\alpha_S^n l^{2n-2}$
respectively. According to this nomenclature, the calculations
\cite{ellis} and \cite{kulesza} are NNLL and NNNLL respectively.

In the $b$-space formalism, following \cite{fnr},
the differential cross section reads:
\begin{eqnarray}
{{d^2\sigma}\over{dm_V^2dq_T^2}} &=& {{\sigma_0}\over{4\pi}}
\sum_{q,q'}|V_{q\bar q'}|^2
\int_0^1{dx_1 dx_2\; \delta(x_1x_2-\tau)}\int{d^2b\; e^{i\vec{q_T}\cdot\vec b}}
\nonumber\\
&\times& \left[ f_{q/1}(x_1,c_1/b)f_{\bar q'/2}(x_2,c_1/b)\exp[S(m_V,b)]+
(q\leftrightarrow\bar q')\right]\;,\end{eqnarray}
where $c_1$ and $c_2$ are integration constants of order 1
and $S(m_V,b)$ is the Sudakov exponent in $b$-space.

For high $b$ values, i.e. small $q_T$, non-perturbative effects are
taken into account via a Gaussian function
$F_{NP}=\exp(-gb^2)$, as suggested in \cite{ly}.
In both \cite{ellis} and \cite{fnr} the value $g=3$ $\mathrm{GeV}^2$ 
is chosen.

Also, in order to allow resummed calculations to be reliable even at large 
$q_T$, we wish to match the calculations of \cite{ellis} and 
\cite{fnr} to the exact ${\cal O}(\alpha_S)$ result.
We add the first-order cross section to the resummed result and, in order to
avoid double counting, we subtract off the term which they have in common,
which is the $q_T\to 0$ limit of the exact
${\cal O}(\alpha_S)$ result. According 
to our prescription, the matching works fine if at the 
point $q_T=m_V$ we have a continuous distribution. 
\section*{Comparison of HERWIG and resummed calculations}
A detailed and general discussion on the comparison of angular-ordered 
parton shower algorithms 
with resummed calculations for Drell--Yan processes 
was already performed in \cite{cmw}, where
the authors showed that, for $\tau\to 1$, 
HERWIG always accounts for the term $A_1$,
corresponding to the leading logarithms in the exponent, and $B_1$ as well.
Furthermore, one is able to account
for the NLL term $A_2$ by simply modifying the Altarelli--Parisi
splitting function introducing a second-order contribution
\begin{equation}  
P_{qq}'(\alpha_S,z)= {{\alpha_S}\over{2\pi}}C_F{{1+z^2}\over{1-z}}
+{{C_F}\over 2}\left( {{\alpha_S}\over{\pi}}\right)^2{K\over{1-z}},
\end{equation}
where the $K$ factor is given by:
\begin{equation}
 K=C_A\left( {{67}\over{18}}-{{\pi^2}\over 6}\right) -{5\over 9}N_f,
\end{equation}
$N_f$ being the number of flavours, $C_F=4/3$ and $C_A=1/2$. 
This is equivalent to redefining the QCD parameter $\Lambda$ to
the `Monte Carlo' $\Lambda_{MC}$: 
\begin{equation}
\Lambda_{MC}=\Lambda\exp(K/4\pi\beta_0),
\end{equation}
with $\beta_0=(11C_A-2N_f)/(12\pi)$.
Even after these replacements, 
the HERWIG algorithm cannot be considered
completely accurate at the next-to-leading level, since it is still
missing higher-order contributions in the strong coupling constant or
the parton distribution functions (see, for instance, the discussion in 
\cite{lh}).

\begin{figure} 
\centerline{\epsfig{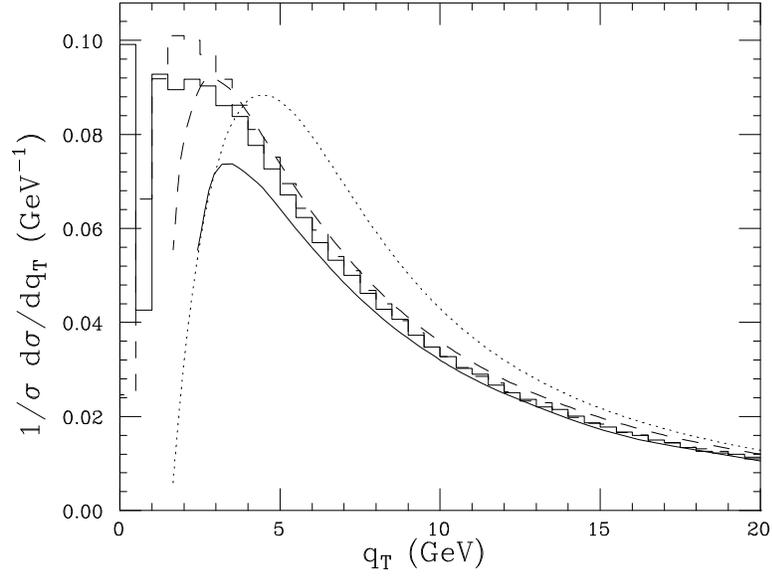}}
\vspace{10pt}
\caption{The $W$ $q_T$ distribution at the Tevatron, according to HERWIG
  with matrix-element corrections, with zero intrinsic $q_T$ (solid
  histogram) and an $q_{T{\mathrm{int}}}$ of 1~GeV (dashed histogram), compared
  with the resummed results of [\ref{fnr}] in $q_T$-space (solid line) and
  in $b$-space (dotted line) and of [\ref{ellis}] (dashed line) 
in $q_T$-space.}
\label{fig8}
\end{figure}

\begin{figure} 
\centerline{\epsfig{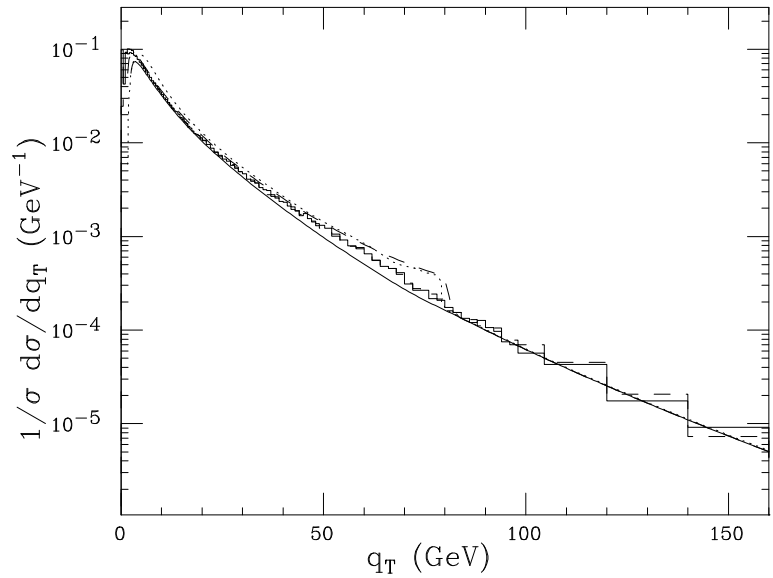}}
\vspace{10pt}
\caption{As in Fig.~\ref{fig8}, but with the resummed results matched
  with the exact ${\cal O}(\alpha_S)$ matrix element.}
\label{fig9}
\end{figure}

In Fig.~\ref{fig8} we show the $W$ transverse momentum distribution 
at the Tevatron in the low-$q_T$ range according to HERWIG 6.1 
and the calculations
of \cite{ellis} in $q_T$-space and of \cite{fnr} in  
$q_T$- and $b$-space.
The HERWIG curve lies within the range of the resummed 
calculations, which is a reasonable result, 
considering that we are actually comparing
different approaches.
In Fig.~\ref{fig9} we consider the whole $q_T$ range, with the resummations
matched to the exact first-order amplitude.
We find that the matching works fine only for  
the approach \cite{fnr} in $q_T$-space, the others showing
a step at $q_T=m_W$, due to uncompensated contributions of order
$\alpha_S^2$ or higher. 
The well-matched distribution agrees with the HERWIG 6.1 prediction at large 
$q_T$.
\section*{Conclusions}
We studied the initial-state radiation in vector boson production according
to the HERWIG event generator and some resummed calculations.
In particular, we investigated the effect of the recently-implemented 
matrix-element corrections to the HERWIG algorithm.
We found a big effect of such corrections on
$W/Z$ transverse momentum distributions at the Tevatron and at the LHC,
and good agreement with the D\O\ and CDF data, with a crucial
role played by such corrections in order to be able to fit in with the
data at large $q_T$. We also found that, even though the spectra at
small $q_T$ do depend on the intrinsic non-perturbative transverse momentum,
the ratio of the $W$ to the $Z$ spectrum is roughly
independent of it.
We then considered some resummed calculations, which we matched to the 
exact ${\cal O}(\alpha_S)$ matrix element, which makes them
reliable at large $q_T$ as well.
We found reasonable agreement of such approaches with HERWIG and
fine matching only for the calculation which keeps all the next-to-leading
logarithms in the Sudakov exponent in $q_T$-space.

Finally, we have to say that the discussed method of improving the 
initial-state radiation in parton-shower Monte Carlo simulations can
be extended to a wide range of interesting processes for the phenomenology
of hadron colliders. The implementation of hard and soft corrections
to top and Higgs production is in progress.
\section*{Acknowledgements}
The presented results have been obtained in collaboration with Mike Seymour.
We also acknowledge Lynne Orr for a careful reading of this manuscript.
This work was supported by grant number DE-FG02-91ER40685 from the U.S.
Dept. of Energy.
 
\end{document}